\documentclass[preprint]{revtex4}

\usepackage[english]{babel}

\usepackage[letterpaper,top=2cm,bottom=2cm,left=3cm,right=3cm,marginparwidth=1.75cm]{geometry}

\usepackage{amsmath}
\usepackage{physics}
\usepackage{graphicx}
\usepackage[colorlinks=true, allcolors=blue]{hyperref}

\begin{document}
\title{Dynamics of non-Gaussian fluctuations in model A}
\author{Thomas Sch\"afer$^1$ and Vladimir Skokov$^{1,2}$}
\affiliation{$^1$ Department of Physics, North Carolina State University,
Raleigh, NC 27695}
\affiliation{$^2$ RIKEN BNL Research Center, Brookhaven National Laboratory, 
Upton, NY 11973, USA}
\begin{abstract}
 Motivated by the experimental search for the QCD critical point
we perform simulations of a stochastic field theory with purely 
relaxational dynamics (model A). We verify the expected dynamic 
scaling of correlation functions. Using a finite size scaling 
analysis we obtain the dynamic critical exponent $z=2.026(56)$. 
We investigate time dependent correlation functions of higher 
moments $M^n(t)$ of the order parameter $M(t)$ for $n=1,2,3,4$. 
We obtain dynamic scaling with the same critical exponent $z$ 
for all $n$, but the relaxation constant depends on $n$. We also 
study the relaxation of $M^n(t)$ after a quench, where the 
simulation is initialized in the high temperature phase, and 
the dynamics is studied at the critical temperature $T_c$. We 
find that the evolution does not follow simple scaling with the 
dynamic exponent $z$, and that it involves an early time rise 
followed by late stage relaxation. 
\end{abstract}
\maketitle

\section{Introduction}
\label{sec:intro}

 Fluctuation observables are an important tool in the experimental search 
for a possible critical endpoint in the QCD phase diagram
\cite{Stephanov:1998dy,Bzdak:2019pkr,Bluhm:2020mpc,An:2021wof}. The main idea 
is that the system created in a heavy ion collision can be described as a 
strongly interacting fluid, which is characterized by a local temperature as 
well as baryon, isospin, and strangeness chemical potentials. In heavy ion
experiments there are a number of control parameters, such as the beam 
energy, the system size, the impact parameter and the rapidity of the 
observed particles, that affect the trajectory of the strongly interacting 
fluid and the location of the freezeout surface in the phase diagram of 
QCD. If the freezeout surface is close to a critical point, then this will 
manifest itself as an enhancement of event-by-event fluctuations in 
a number of observables, for example in the variance of the number 
of net protons. It was recognized  that it may be advantageous to consider
higher order cumulants, such as the skewness and kurtosis
\cite{Ejiri:2005wq,Stephanov:2008qz,Asakawa:2009aj,Stephanov:2011pb,Friman:2011pf}, 
because these observables show a stronger dependence on the correlation length 
near the critical point, and they may exhibit a non-monotonic behavior as a 
function of the beam energy. 

 A heavy ion collision is a short-lived, very dynamic, event. This implies that 
non-equilibrium phenomena, such as critical slowing down and memory effects cannot
be ignored \cite{Berdnikov:1999ph,Nahrgang:2018afz,Akamatsu:2018vjr,Bluhm:2020mpc,
An:2021wof}. In terms of the experimental observation of critical behavior this 
fact may indeed be beneficial, as memory effects imply that fluctuations 
observables at freezeout encode the possible transit of a critical point 
earlier during the evolution. However, non-equilibrium effects also complicate 
the interpretation of experimental results, because critical slowing down limits 
the overall magnitude of critical effects, and non-equilibrium phenomena affect 
individual observables in different ways. As a consequence, we may not be 
able to map different cumulants at freezeout onto a single point in the 
phase diagram. 

 Previous work on this issue has mostly focused on the time evolution of the 
two-point function \cite{Stephanov:2017ghc,Akamatsu:2018vjr}. Notable exceptions
include the work in \cite{Mukherjee:2015swa,Nahrgang:2018afz,An:2020vri,
Sogabe:2021svv}, but these studies use approximations to truncate the 
hierarchy of correlation functions, or solve the dynamics in a restricted
geometry. Our goal in this work is to study the time evolution and equilibration
time of non-Gaussian fluctuations in a 3+1 dimensional framework. For simplicity 
we will consider the critical behavior of a non-conserved order parameter in a
relaxational theory without couplings to a conserved energy or momentum density.
This is known as model A in the classification of Hohenberg and Halperin
\cite{Hohenberg:1977ym}. A possible critical endpoint in the QCD phase diagram 
is believed to be in the universality class of model H \cite{Son:2004iv}, and the
chiral critical behavior in the vicinity of a second order chiral phase transition
is in the universality class of model G \cite{Rajagopal:1992qz,Nakano:2011re,
Florio:2021jlx}. In the present work we study the dynamics of model A on a 
three dimensional lattice. Studies of kinetic Ising models are reported in
\cite{Hasenbusch:2019gmx} (and references therein) and relativistic stochastic
models have been investigated in \cite{Schweitzer:2020noq,Schweitzer:2021iqk}.
  
\section{Model A}
\label{sec:modA}

 For a single component field the dissipative dynamics is described by 
\begin{align}
\label{modA}
    \frac{\partial \phi(t, \vec{x})}{\partial t} = 
    - \Gamma\,  \frac {\delta {\cal H}} {\delta \phi(t, \vec{x})} 
    + \zeta (t, \vec{x}), 
\end{align}
where $\Gamma$ is the relaxation rate and ${\cal H}$ is the Hamiltonian
\begin{align}
\label{H_Ising}
    {\cal H}  = \int d^dx \left[ \frac{1}{2} (\nabla \phi)^2  
    +  \frac{1}{2} m_0^2 \phi^2(t,\vec{x}) 
    +   \frac{1}{4} \lambda  \phi^4(t,\vec{x})  
    - h(t,\vec{x}) \phi (t,\vec{x})\right] \, ,
\end{align}
and $h$ is an external field. The noise term $\zeta (t, \vec{x})$ describes 
the interaction with microscopic degrees of freedom. Usually it is modelled 
as a random field with zero mean and white noise spectrum  
\begin{align}
    \langle \zeta (t, \vec{x}) \zeta (t', \vec{x}') \rangle = 
    2 T\, \Gamma\, \delta(\vec{x}-\vec{x}')\delta(t-t')\, .
\end{align}
In this work we will discretize the Hamiltonian on a spatial lattice 
with lattice spacing $a$, adopting units so that $a=1$. We use periodic boundary conditions in our simulations. 
We write the lattice field as $\phi(t,\vec{x}) \to \phi(\vec{x})$, 
where we have dropped the explicit dependence on $t$ to simplify the 
notation, and $\vec{x}=a\vec{n}$ with $\vec{n}$ a vector with integer
components. We have
\begin{align}
    {\cal H}  = \sum_{\vec{x}} \left[ \frac{1}{2}  
    \sum_{\mu=1}^d   (\phi(\vec{x}+\hat{\mu}) - \phi(\vec{x}) )^2  
    +  \frac{1}{2} m_0^2 \phi^2(\vec{x}) 
    +   \frac{1}{4} \lambda  \phi^4(\vec{x})  - h \phi (\vec{x})\right] \, . 
\end{align}
The vector $\hat \mu$ has a unit size and only one non-zero entry at 
$\mu$-th position. E.g. for $\mu=2$ and $d=3$, $\hat \mu = (0,1,0)$. 
In the following we will need the change in the Hamiltonian as we make 
a local update of the field. Changing the field $\phi_{\rm old} \to 
\phi_{\rm new}$ at a fixed position $\vec{x}$ leads to 
\begin{align}
    \Delta {\cal H}  &= 
    d (\phi^2_{\rm new}(\vec{x}) - \phi^2_{\rm old}(\vec{x}))
    -  (\phi_{\rm new}(\vec{x}) - \phi_{\rm old}(\vec{x}))
    \sum_{\mu=1}^d \left( \phi(\vec{x}+\hat{\mu}) + \phi(\vec{x}-\hat{\mu})  
    \right)\notag \\   
    &+\frac{1}{2} m_0^2 ( \phi^2_{\rm new}(\vec{x}) - \phi^2_{\rm old}(\vec{x}))  
    +   \frac{1}{4} \lambda  ( \phi^4_{\rm new}(\vec{x}) 
    - \phi^4_{\rm old}(\vec{x}))  - h (\phi_{\rm new}(\vec{x}) 
    - \phi_{\rm old}(\vec{x})) \,.
\end{align}
Following the recent work of Florio et al.~\cite{Florio:2021jlx} we will
simulate the stochastic evolution using a Metropolis algorithm. Similar 
methods have been employed in the Langevin simulation described in 
\cite{Moore:1998zk} and in kinetic Ising models \cite{Hasenbusch:2019gmx}.
The time-dynamics is modelled using a fixed time step size $\Delta t$. We sweep 
through the lattice using a checkerboard pattern. At every site $\vec{x}$ we 
perform a trial update
\begin{align}
    \phi_{\rm new}(\vec{x}) 
    =  \phi_{\rm old}(\vec{x}) + \sqrt{2 \Delta t  \Gamma} \xi\,.  
\end{align}
Here $\xi$ is a random number sampled from a Gaussian distribution with zero 
mean and unit variance. The trial update is accepted, that is 
\begin{align}
    \phi(t+\Delta t, \vec{x}) =  \phi_{\rm new}(\vec{x})\,   
\end{align}
with probability $P={\rm min} (1, e^{-\Delta {\cal H}})$. If the trial update 
is rejected, the field at this lattice site remains unchanged 
\begin{align}
    \phi(t+\Delta t, \vec{x}) =  \phi(t, \vec{x})\,.    
\end{align}
Note that for the purely relaxational model studied here the relaxation
rate sets the unit of time. For simplicity we set $\Gamma=1$, and time
is measured in units of the inverse relaxation constant.

\section{Statics}
\label{sec:statics}

  The time evolution of a generic initial condition via the stochastic
differential equation (\ref{modA}) will generate an ensemble characterized
by the probability distribution $P(\phi)\sim\exp(-{\cal H}/T)$. This 
ensemble can be used to study static properties of the Ising Hamiltonian 
in equ.~(\ref{H_Ising}). As a first step, we determine the critical 
value $m_c^2<0$ of the mass parameter for a given value of $\lambda$.
Here, we have chosen $\lambda = 4$. 

\begin{figure}
\centering
\includegraphics[width=0.65\linewidth]{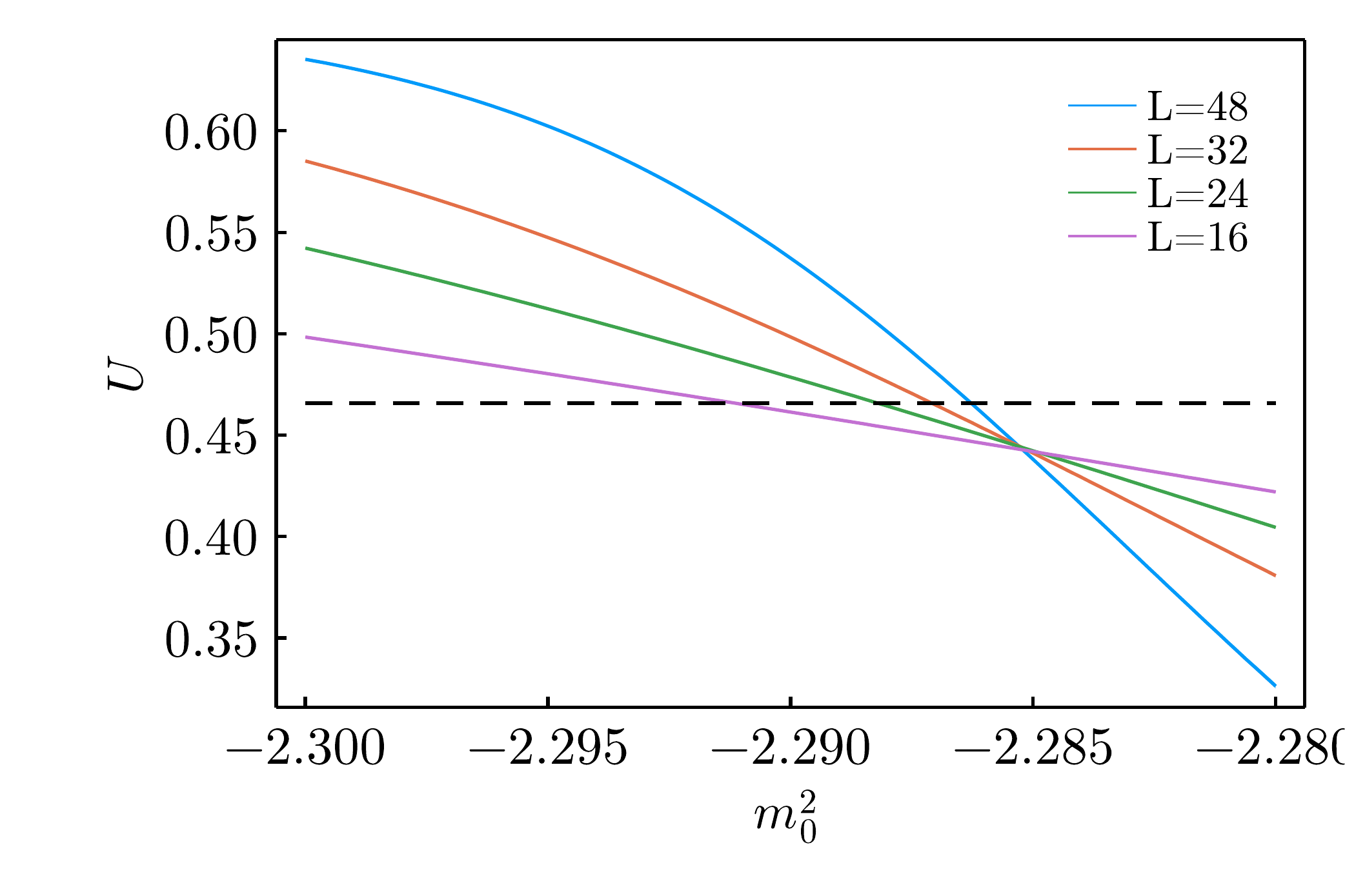}
\caption{\label{fig:Binder}
The Binder cumulant as a function of the control parameter $m_0^2$ for different 
values of the box size $L$. The dashed line show the universal value of $U$
in the infinite volume limit, determined in \cite{Hasenbusch:1998ve}. An 
extrapolation of the location of the crossing of our Monte Carlo results 
with the dashed line to $L\to \infty$ gives $m_c^2 = -2.28587(7)$.
}
\end{figure}

 In order to locate $m_c^2$ from simulations in a finite volume we
use the Binder cumulant method. We define the magnetization 
\begin{align}
\label{M}
    M (t) = \frac{1}{L^3} \sum_{\vec{x}} \phi(t, \vec{x}) 
\end{align}
and compute the time average
\begin{align}
    \langle  O \rangle  = 
    \frac{\sum_t O(t)}{\sum_t 1}. 
\end{align}
Here, the sum should only be taken over configurations that are 
generated after the initial equilibration time, and that are separated
by more than the autocorrelation time. In practice, we perform 
measurements every 100 Monte Carlo sweeps through the entire lattice. 
Near the critical value $m_c^2$ fluctuation observables such as 
$\langle M^2\rangle$ and $\langle M^4\rangle$ show peaks, but in 
a finite volume simulation the location of these peaks is different
from the infinite volume value of $m_c^2$. Binder observed that at
the true critical point $m_c^2$ the leading finite volume corrections
to the Binder cumulant 
\begin{align}
\label{Binder}
    U  = 1 - \frac{ \langle M^4 \rangle } { 3 \langle M^2 \rangle^2  } 
\end{align}
cancel \cite{Binder:1981}. This means that we can use finite volume calculations 
of $U$ to accurately locate the critical point of the infinite system. 

 In practice we perform a series of simulations on a coarse grid in 
$m_0^2$, and then use reweighting to more accurately determine the critical
value of $m_0^2$. Reweighting for a given observable $O[M]$ is performed around ${\bar m}_0^2$ via 
\begin{align}
\langle O[M] \rangle_{m_0^2}  = 
\frac{ \sum_t e^{ - \frac{\delta m^2}{2} \sum_{\vec {x}}  \phi^2(t,\vec{x}) } 
          O[M(t)]  } 
 {\sum_t e^{ - \frac{\delta m^2}{2} \sum_{\vec {x}}  \phi^2(t,\vec{x}) }  }\,.
\end{align}
Here $\delta m^2 = m_0^2 - {\bar m}^2_{0}$ and the field configurations are sampled at  ${\bar m}_0^2$.    Therefore in the code one can simply save 
$M(t)$  and $\sum_{\vec {x}}  \phi^2(t,\vec{x}) $ for many $t$; the analysis 
is then can be executed separately and independently for different target $m_0^2$. Large
values of $|\delta m^2|$ (even of order 0.1) are not only problematic because of 
reweighting but also due to large values of    the sum $\sum_{\vec {x}}  
\phi^2(t,\vec{x})$ and the corresponding underflow/overflow while evaluating the 
exponent. Instead one can simply compute 
\begin{align}
 \langle O[M] \rangle_{m_0^2}  = 
    \frac{ \sum_t e^{ - \frac{\delta m^2}{2} 
      ( \sum_{\vec {x}}   \phi^2(t,\vec{x}) - c) }    O[M(t)]  } 
       {\sum_t e^{ - \frac{\delta m^2}{2} 
       (\sum_{\vec {x}}  \phi^2(t,\vec{x}) - c) }  }
\end{align}
where $c$ is a constant defined by 
\begin{align}
    c = \left\langle \sum_{\vec {x}}   \phi^2(t,\vec{x})  \right\rangle_{{\bar m}_0^2} \,.
\end{align}
The approximate value of critical $m_c^2$ was determined by running the 
simulations on coarse lattices, see Fig.~\ref{fig:Binder}. Then longer 
simulation were done at single value of the parameter ${\bar m}_0^2 = -2.288$, 
and reweighting technique was used to recover the dependence on  $m_0^2$ in 
the vicinity of ${\bar m}_0^2$. 

\begin{figure}[t]
\centering
\includegraphics[width=0.49\textwidth]{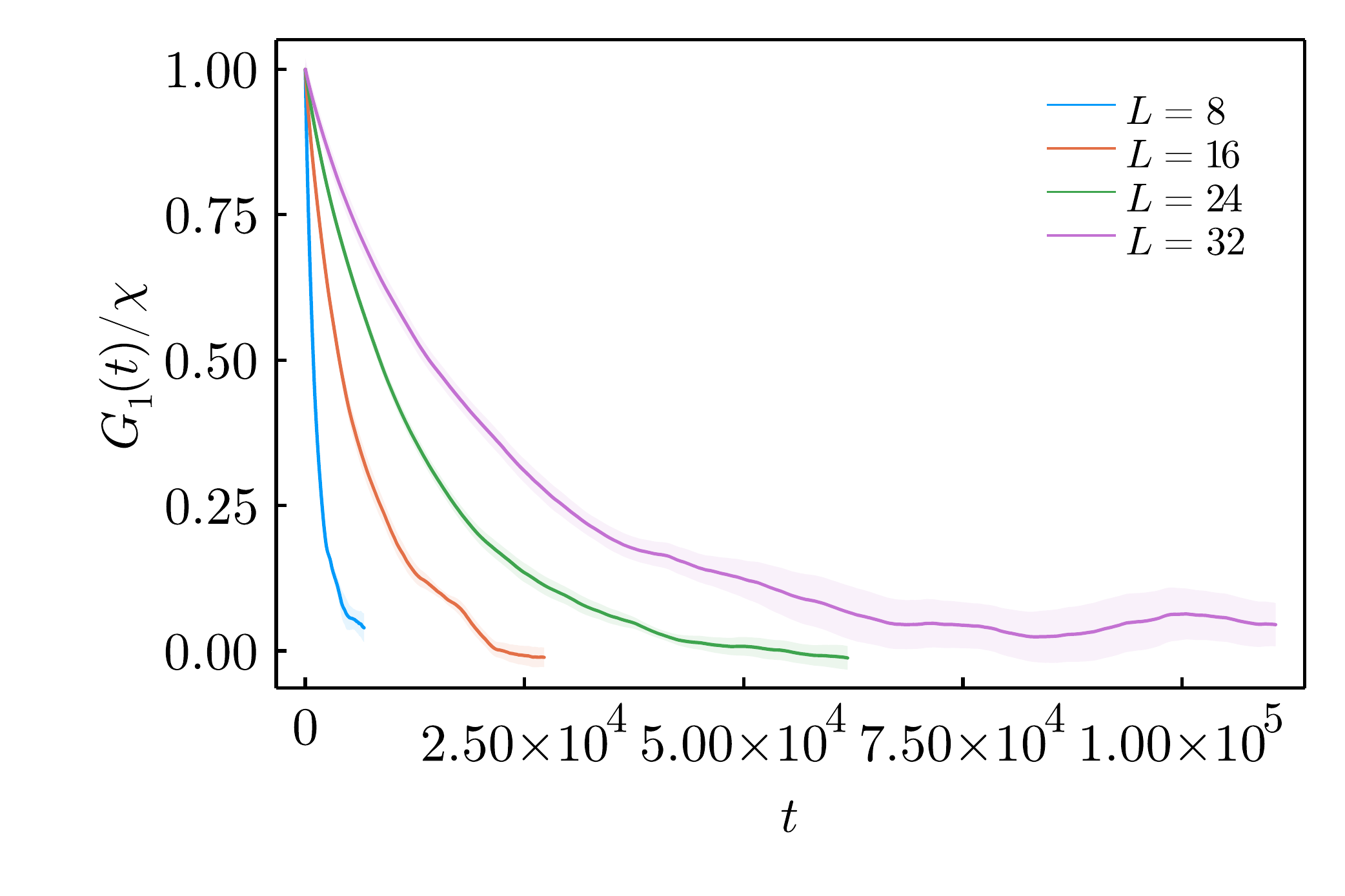}
\includegraphics[width=0.49\textwidth]{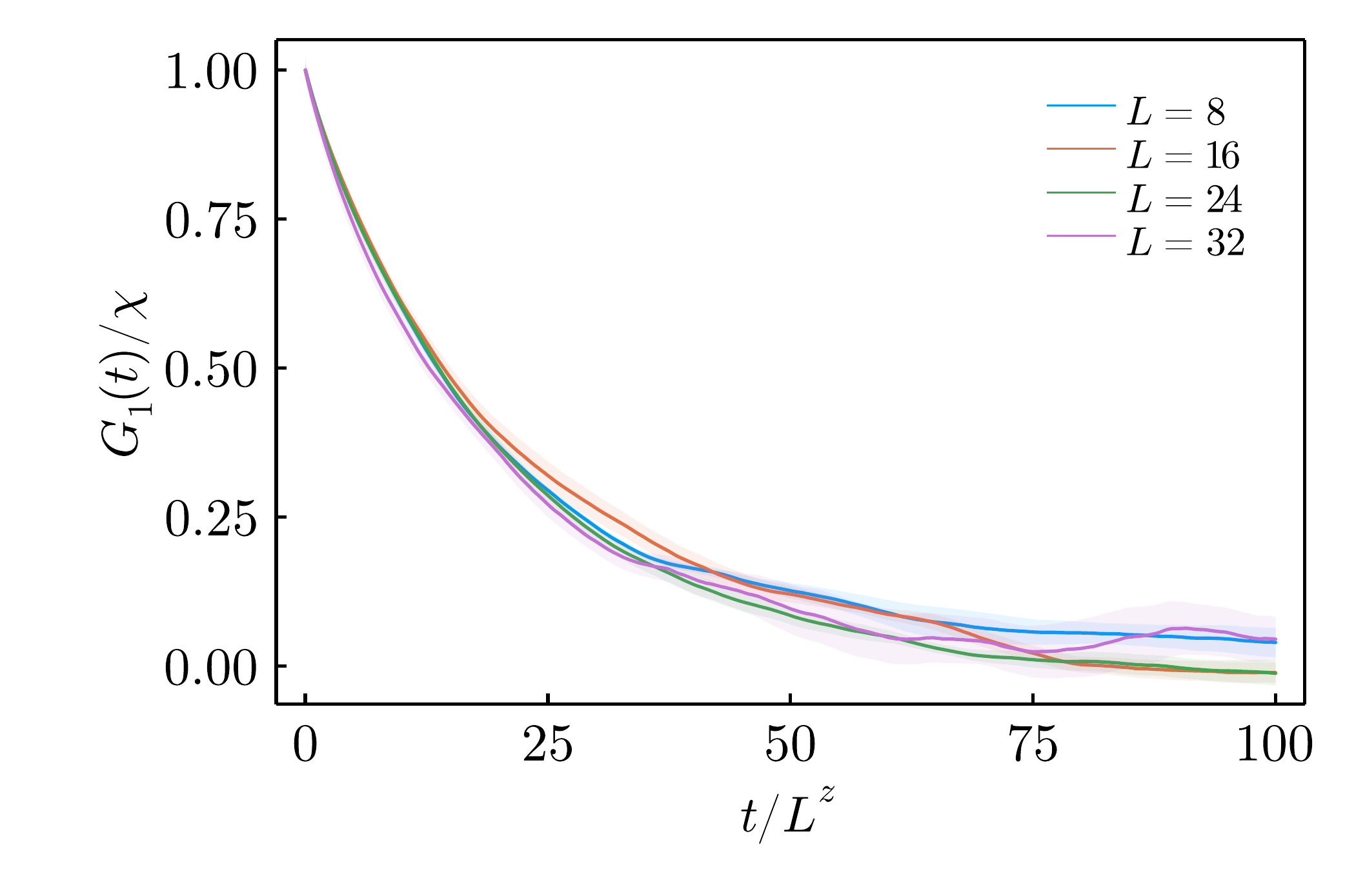}
\caption{Correlation function $G_1(t)/\chi$ as a function of time for 
various system sizes at the critical point. Left panel: Correlation functions 
as a function of time $t$. Right panel: Correlation functions as a function 
of the a function of the scaled time variable $t/L^z$.
\label{fig:C_tau}}
\end{figure}

 The crossing points between the solid lines in Fig.~\ref{fig:Binder} indicate 
that $m_c^2\simeq -2.285$. A very precise value of $m_c^2$ can be obtained using 
the strategy adopted in \cite{Florio:2021jlx}. We make use of the universal value 
of the Binder cumulant at the critical point, $U_c = 0.4658$ \cite{Hasenbusch:1998ve}.
This value is shown as the dashed line in Fig.~\ref{fig:Binder}. Then we find
the location $m_{\cross}^2(L)$ at which the Binder cumulant $U(m^2,L)$ intersects
the line $U=U_c$. Finite size scaling predicts $m_{\cross}^2 (L) = m_c^2 + 
C L ^ {- 1/\nu-\omega}$.  Using the values of the critical exponents $\omega = 
0.8303$ and $\nu = 0.62999(5)$ from \cite{El-Showk:2014dwa} to fit the data for 
$L=24$, 32, 48 we get $m_c^2 = -2.28587(7)$. 

\section{Dynamics}
\label{sec:dynamics}

\begin{figure}
\centering
\includegraphics[width=0.65\textwidth]{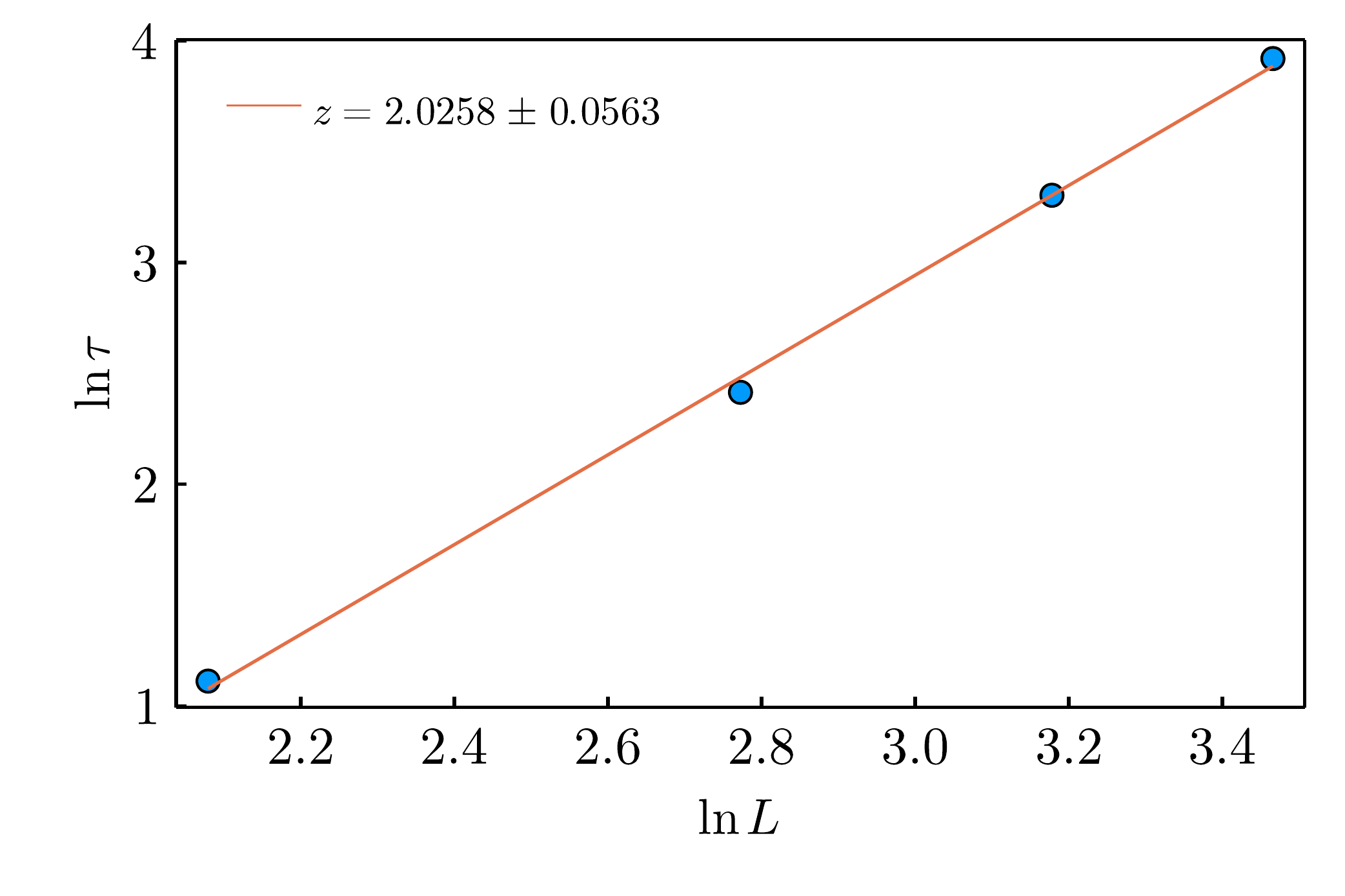}
\caption{Relation between the decay time $\tau$ extracted from the 
correlation function $G_1(t)$ and the linear box size $L$. The line 
shows the best fit with $z=2.0258$.
\label{fig:z_fit}}
\end{figure}

 An important feature of the dynamics near the critical point is dynamical
scaling. It implies that near the critical point the correlation function 
$G_1(t,x)=\langle \phi(0,0)\phi(t,x)\rangle$ behaves as $G_1(t,x)\sim 
f(t/\tau,x/\xi)$, where $f$ is a universal function, $\xi$ is the correlation
length, and $\tau\sim \xi^z$ is the correlation time. The quantity $z$ is known 
as the dynamic critical exponent. The dynamic exponent can be determined by 
studying the scaling of the relaxation time of a mode with wave number $k\sim
\xi^{-1}$ as a function of the correlation length. Here, we use a simpler method,
which is based on finite size scaling \cite{Zinn-Justin:2002ecy}. At zero external 
magnetic field, the correlation time  in a finite volume of linear size is
\begin{align}
\label{tau-scal}
    \tau (T, L)  = L^z f_\tau \left( \Delta T L^{1/\nu}  \right)\,. 
\end{align}
We work sufficiently close to the critical point to approximate  
$f \left( \Delta T L^{1/\nu}  \right) $ by $f \left( 0 \right) $. In the case of
model A, the prediction of the $\varepsilon$ expansion is 
\cite{Zinn-Justin:2002ecy,Antonov:1984,Folk:2006ve}
\begin{align}
\label{z-eps}
    z = 2 + R \, \eta, \quad R = \left( 6 \ln \frac43 -1  \right) 
      \left(1 - \varepsilon\cdot 0.1885 + ...  \right)
\end{align}
with $\eta=3\varepsilon^2/162$. The conformal bootstrap gives $\eta \approx 0.0363$ 
\cite{Alday:2015ota}. Using the latter value we obtain  $z\simeq 2.02$. Dynamic scaling
is expected to extend to higher $n$-point functions \cite{Zinn-Justin:2002ecy}, and
to correlation functions of higher dimension operators. In the following we will 
consider the correlation functions
\begin{align}
    G_n(t) = \langle M^n(0)M^n(t) \rangle , 
\end{align}
where $M(t)$ is the magnetization defined in equ.~(\ref{M}), and for even 
$n$ we subtract the asymptotic value, $G_n(t)\to G_n(t)-\langle M^n(0)
\rangle^2$. Note that $G_1(t)$ is the volume integral of the function 
$G_1(t,x)$ defined above. 

\begin{figure}[t]
\centering
\includegraphics[width=0.49\textwidth]{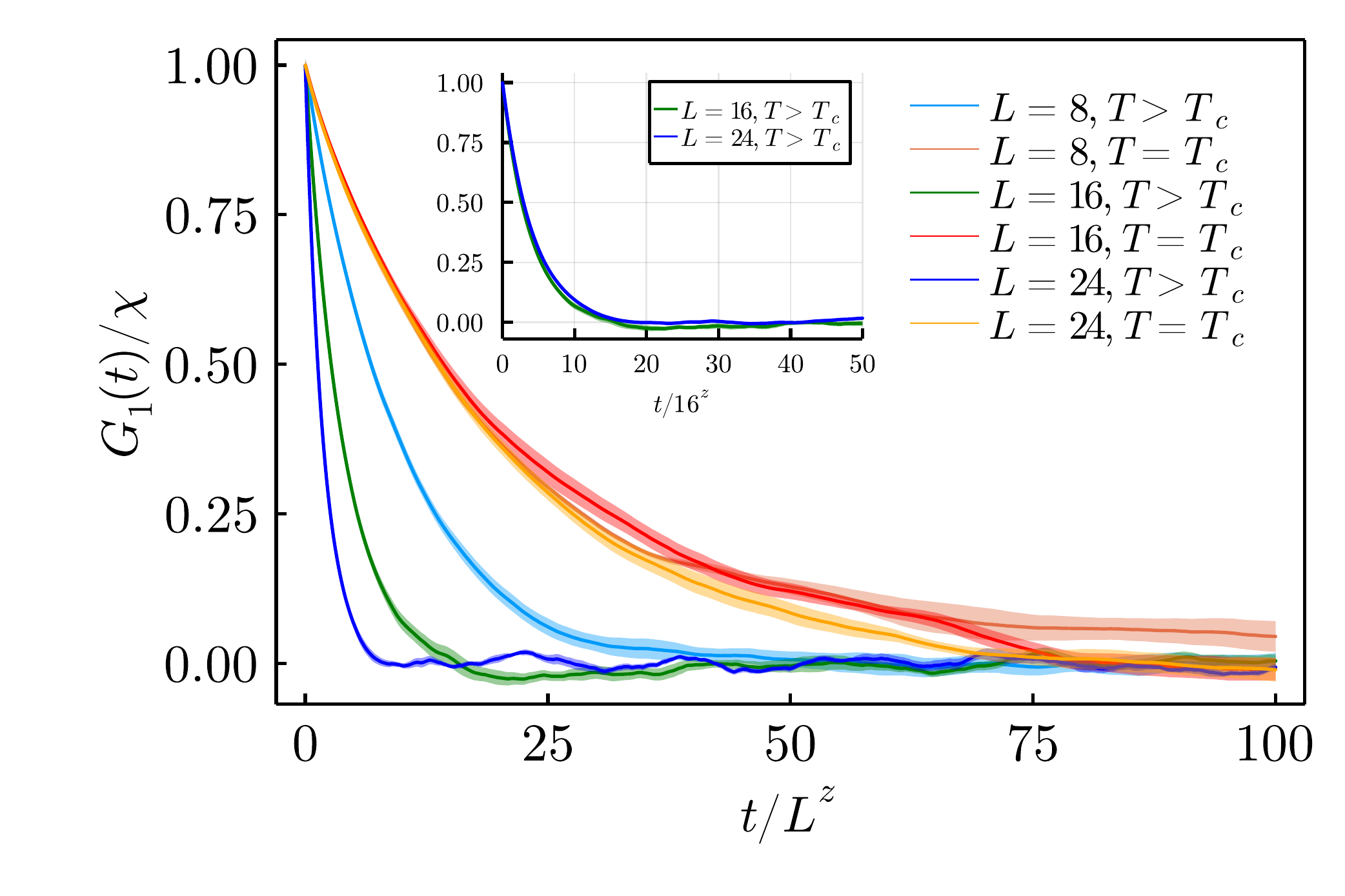}
\includegraphics[width=0.49\textwidth]{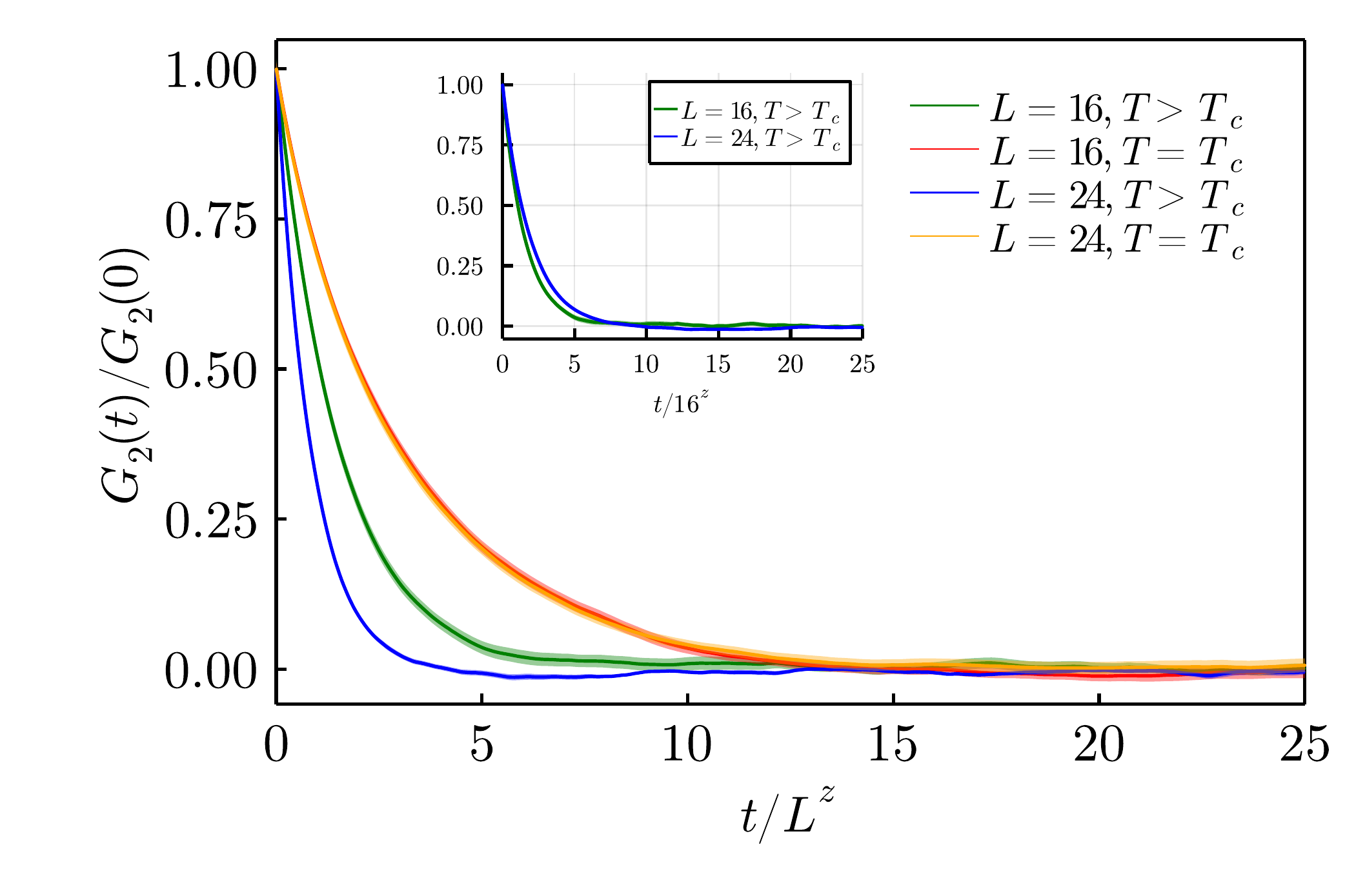}
\caption{$G_1(t)/\chi$ (left panel) and $G_2(t)/G_2(0)$ (right panel)
as a function of time for various system sizes $L$ at and above the 
critical point. The figure shows time in units of $L^z$, and the inset
displays the correlation function with no scaling applied.}
\label{fig:abovetc}
\end{figure}

  In Fig.~\ref{fig:C_tau} we show the two-point function of the magnetization
$G_1(t)$ for different volumes at zero external field and directly at the 
critical point $m_c^2$. In the right panel we have scaled the time $t$ by a 
factor $L^z$, using the theory prediction $z=2.02$. We observe that the data
collapse to a universal function. In Fig.~\ref{fig:z_fit} we show the 
scaling of the fitted decay constant $\tau$ with the box size $L$. The 
best fit value of the dynamic exponent is $z = 2.0258\pm 0.0563$, in good
agreement with the theoretical expectation. We also obtain $\ln f_\tau(0) = -3.13 
\pm 0.16$, where $f_\tau(x)$ is defined in equ.~(\ref{tau-scal}).

 In Fig.~\ref{fig:abovetc} we show that dynamical scaling does not hold
away from the critical point. The left and right panel show the two-point
functions of $M$ and $M^2$, $G_1(t)$ and $G_2(t)$, normalized to their 
value at $t=0$. The $T>T_c$ data were taken at $m_0^2= -2.20$. 
The main figure show the correlation function scaled by $L^z$, and the 
inset shows unscaled correlation functions. We observe that both $G_1$
and $G_2$ exhibit dynamic scaling at $T_c$, but that for $T>T_c$ the 
correlation functions are essentially independent of volume. 

\begin{figure}[t]
\centering
\includegraphics[width=0.65\textwidth]{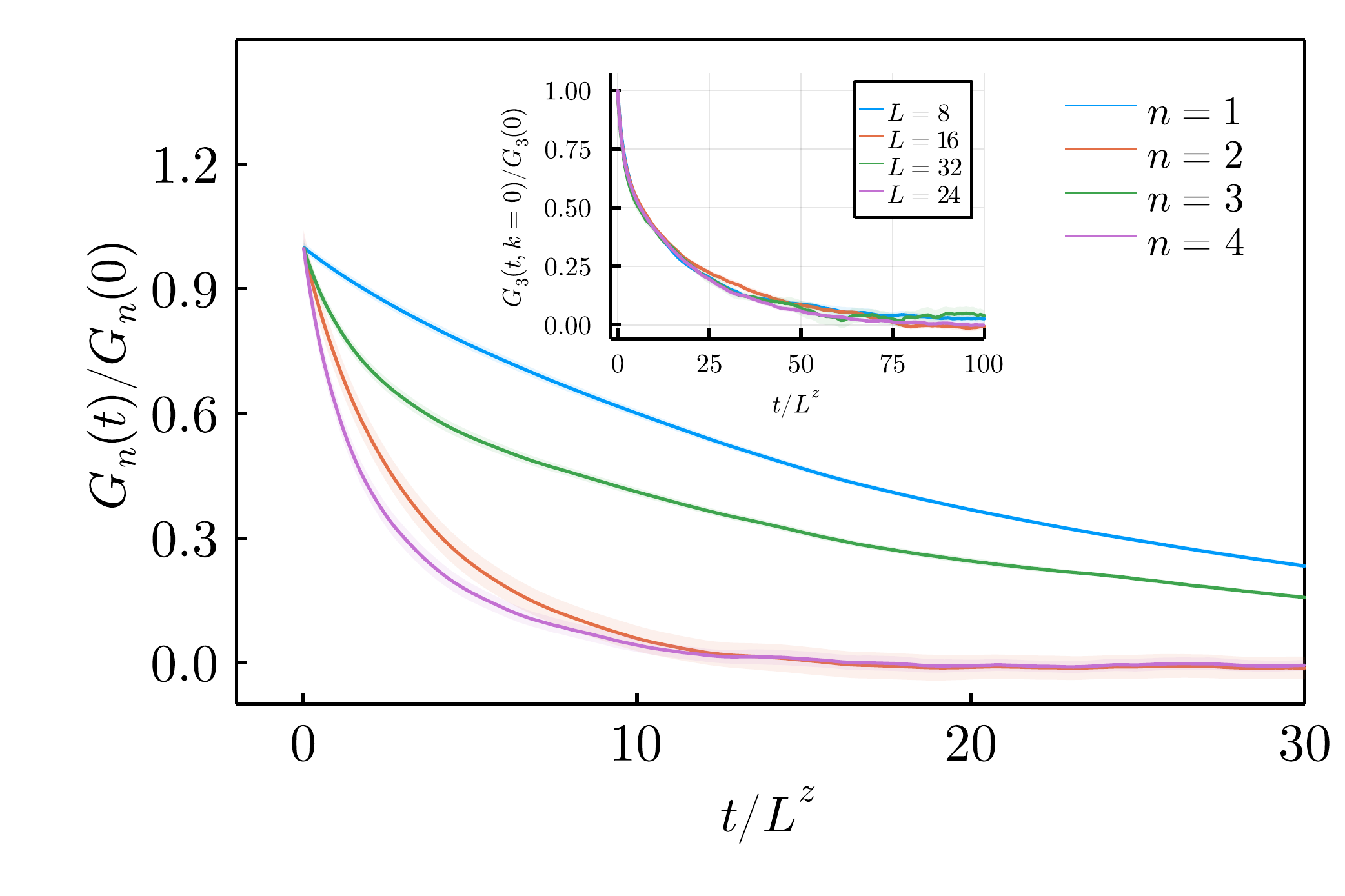}
\caption{$G_n(t)/G_n(0)$ as a function of time $t$ for a fixed 
system size $L$ at the critical point. The inset demonstrates that 
dynamic scaling is present for higher order correlations functions. 
Here, we choose $G_3(t)$ as an illustration, and consider several 
different linear box sizes $L=8,16,24,32$. 
}
\label{fig:higherorder}
\end{figure}

\begin{figure}[t]
\centering
\includegraphics[width=0.49\textwidth]{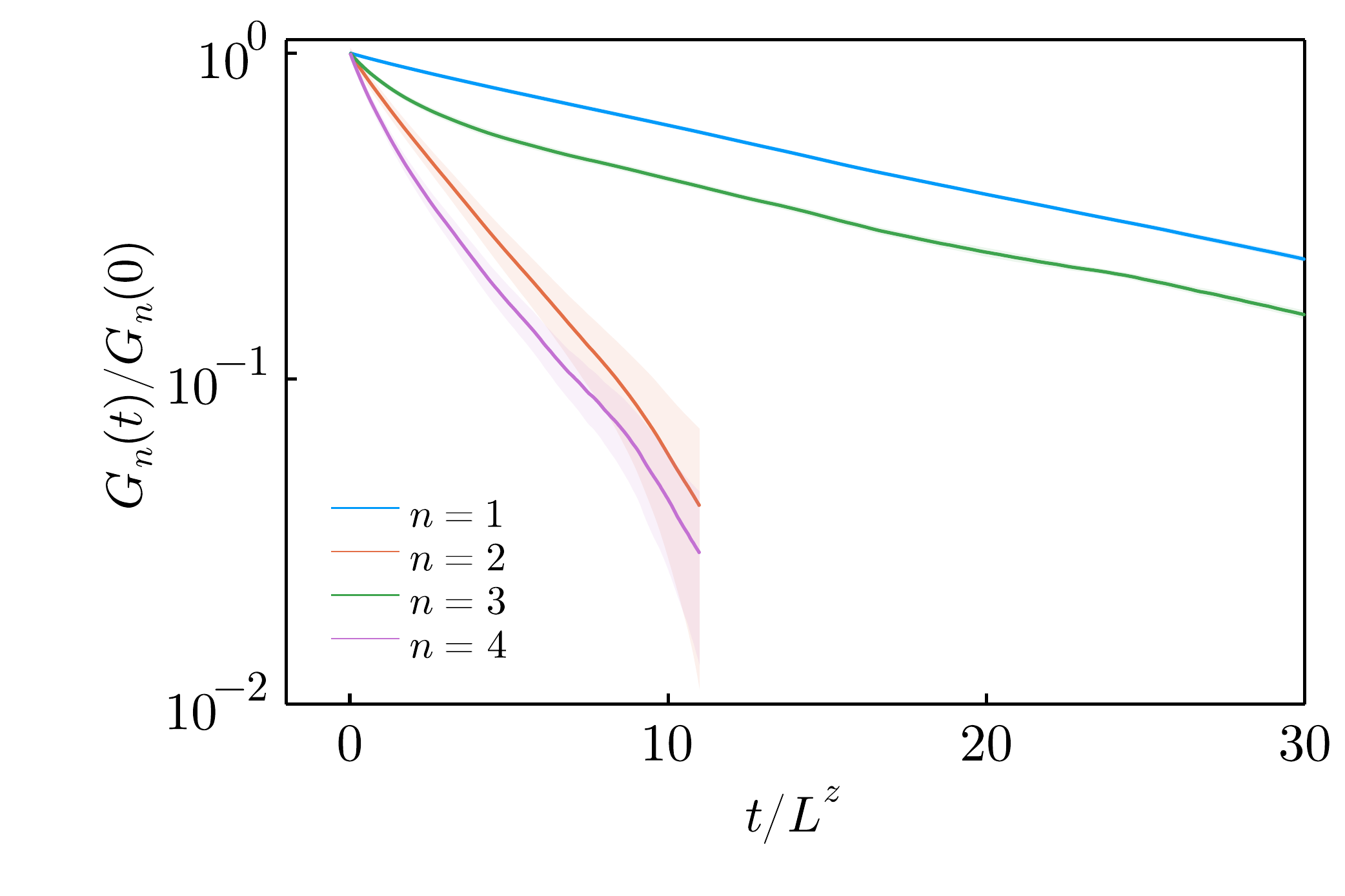}
\includegraphics[width=0.49\textwidth]{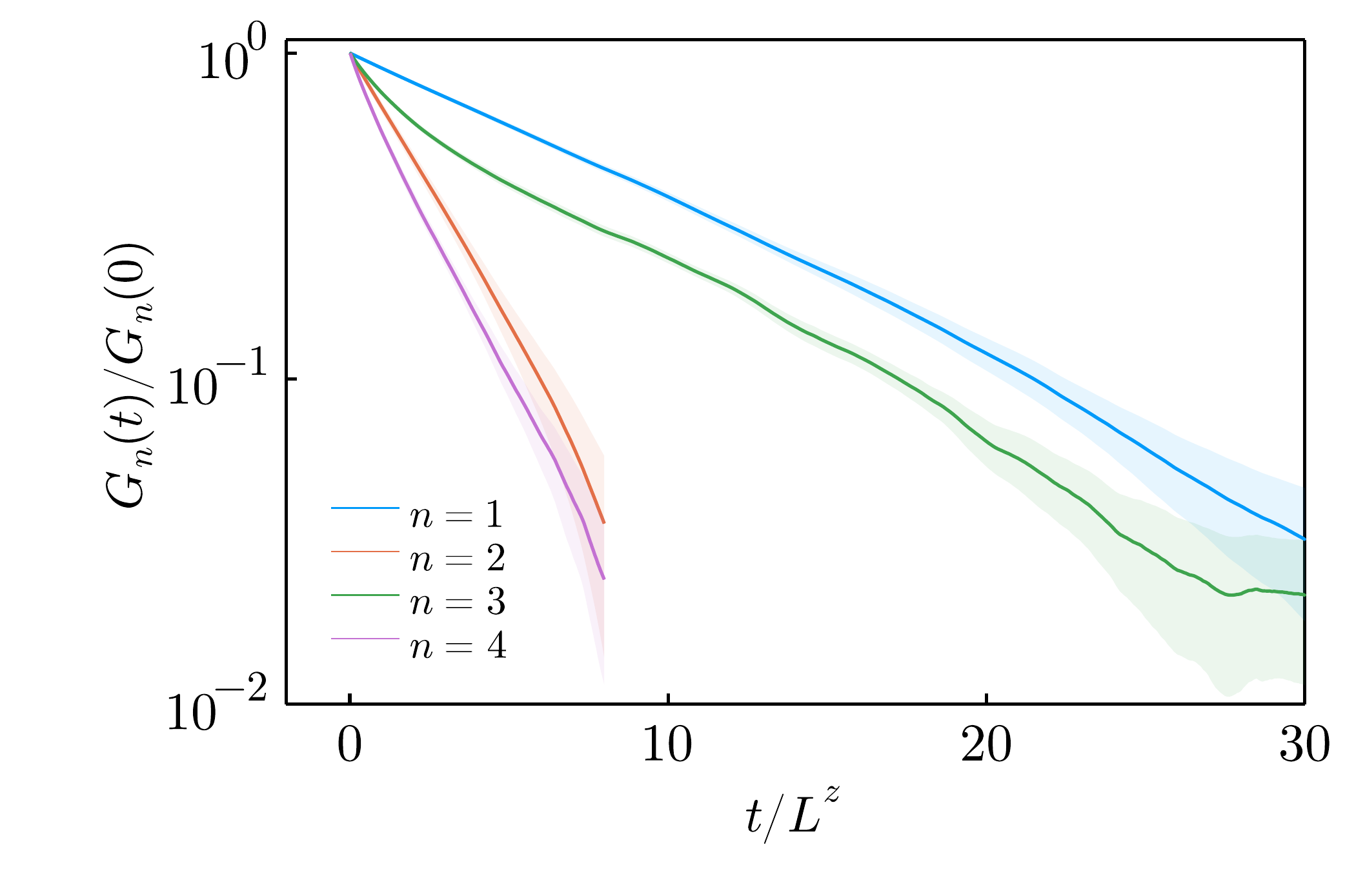}
\caption{
Left panel: This figure shows the same correlation functions as the 
main plot in Fig.~\ref{fig:higherorder}, but displayed on a logarithmic scale. 
The curves for $n=2,4$ are terminated when the lower error band crosses zero.
Right panel: Same set of correlation functions as in the left panel for 
$T>T_c$.}
\label{fig:higherorder_log}
\end{figure}

 The correlation functions of $M^n(t)$ for $n=1,2,3,4$ are shown in  
Fig.~\ref{fig:higherorder} and Fig.~\ref{fig:higherorder_log}. The inset of 
Fig.~\ref{fig:higherorder} shows that dynamic scaling continues to hold. Here, 
we focus on $G_3(t)$ and show that data collapse takes place when $t$ is scaled 
by $L^z$. Similar results hold for $G_2(t)$ and $G_4(t)$. We observe that there 
is a pattern where correlation functions of even powers of $M(t)$ decay
at a similar rate, and the same is true for correlation functions of odd 
powers of $M(t)$. Fig.~\ref{fig:higherorder_log} shows that this pattern holds 
both at $T_c$ and above $T_c$. Note that dynamic scaling is broken at $T>T_c$.
The observed pattern is consistent with the existence of partially disconnected
contributions $G_3(t)\sim \langle M^2\rangle^2 \langle M(0)M(t)\rangle$ and
$G_4(t)\sim \langle M^2\rangle^2 \langle M^2(0)^2M(t)\rangle$. Finally, we 
observe that the asymptotic decay of $G_2(t)$ and $G_4(t)$ is significantly
faster than that of $G_1(t)$ and $G_3(t)$. An exponential fit of the 
asymptotic decay rate indicates that $G_2(t)$ decays about five times 
faster than $G_1(t).$

\begin{figure}[t]
\centering
\includegraphics[width=0.49\textwidth]{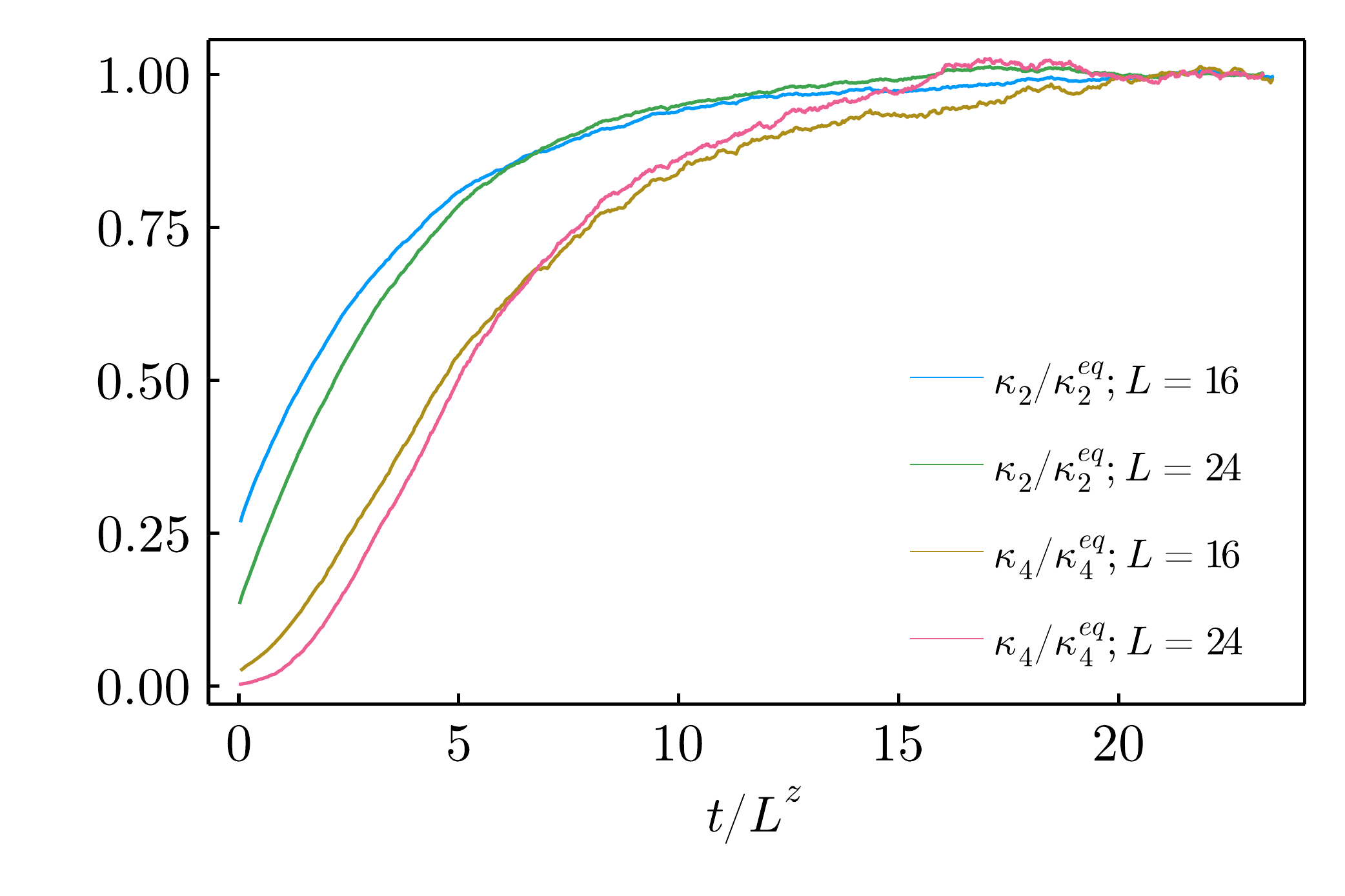}
\includegraphics[width=0.49\textwidth]{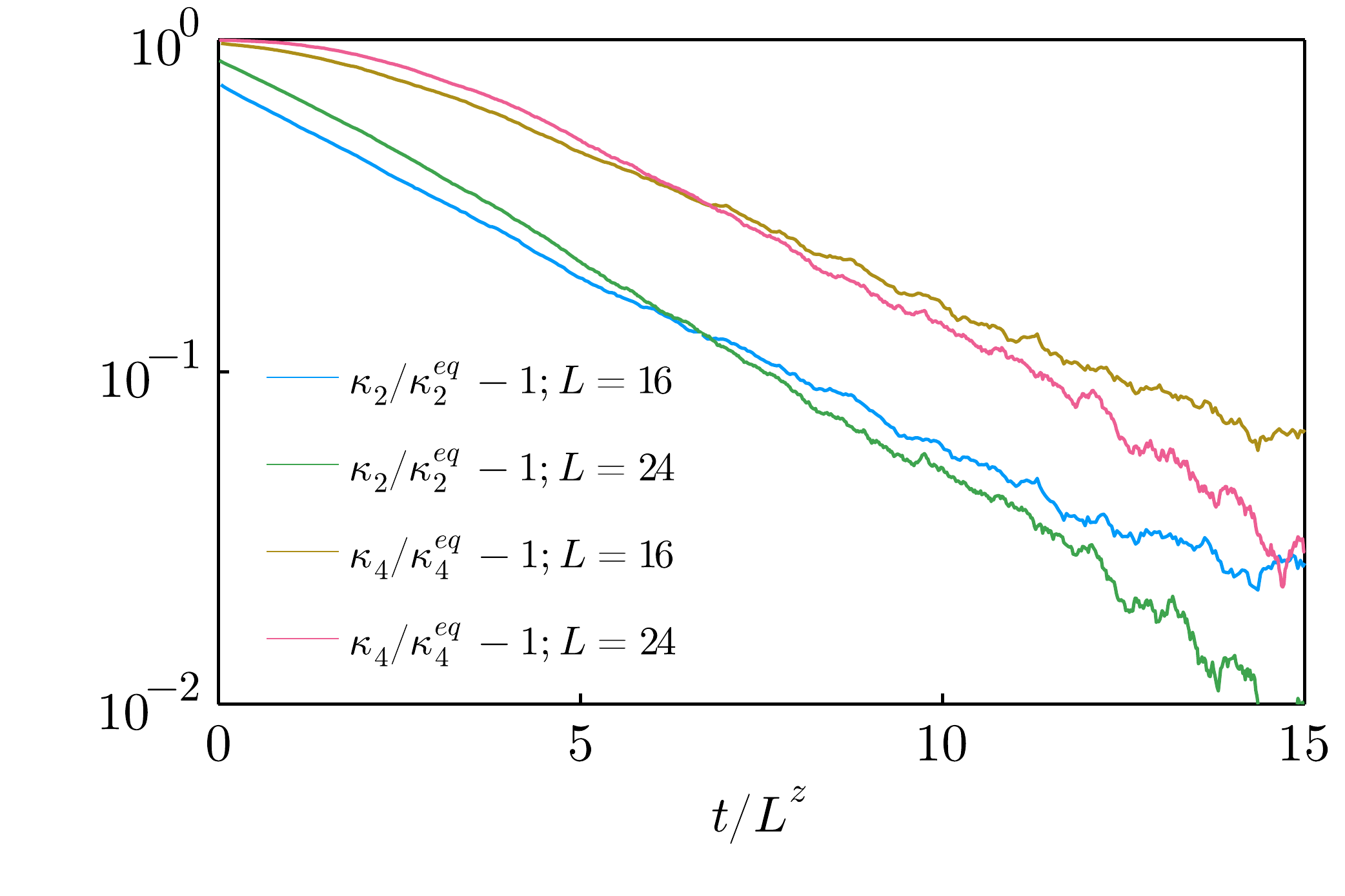}
\caption{
Left panel: Time dependence of the moments $M^2(t)$ and $M^4(t)$ of the 
order parameter after a quench from the high temperature phase to the 
the critical point. The curves are normalized to the asymptotic 
expectation values $\langle M^2\rangle$ and $\langle M^4\rangle$ at the 
critical temperature. We show results for two different linear box sizes,
$L=16$ and $L=24$. 
Right panel: Same functions as in the left panel, but displayed on a 
logarithmic scale and with the asymptotic value subtracted. }

\label{fig:quench}
\end{figure}

 We have also studied the relaxation of $M^n(t)$ after a quench. In 
Fig.~\ref{fig:quench} we show $\langle M^n(t)\rangle$ for an ensemble of 
stochastic evolutions. The initial state is drawn from an ensemble generated
at $T>T_c$, using $m_0^2=-2.20$ as in Fig.~\ref{fig:abovetc} and the right panel
of Fig.~\ref{fig:higherorder_log}. This state is evolved using the dynamic 
equations at the critical point. We observe that the evolution consists of 
at least two stages, an early time rise and a late stage relaxation. Note 
that the early time behavior persists in a regime $t\in[0,t_s]$, where 
$t_s\gg\Gamma^{-1}$ is a macroscopic time scale. Also note that the early 
time dynamics violates dynamic scaling. There is no data collapse if the 
time evolution in different volumes is scaled by $L^z$. Finally, we observe 
that both the initial evolution and the final relaxation of $M^4(t)$ is 
slower than that of $M^2(t)$.

  Breaking of dynamic scaling in the early time behavior of the order 
parameter after a quench to the critical regime was previously studied
in \cite{Janssen:1989,Ritschel:1995,Calabrese:2006ng}. These papers 
characterize the early time rise of the order parameter in terms of 
a slip exponent $\Theta$, which is distinct from the dynamic exponent 
$z$. They also note that the crossover between early time evolution and
late time relaxation takes place at a macroscopic time that depends
on the size of the system.

\section{Conclusions and Outlook}
\label{sec:sum}

 We have studied the time evolution of higher moments of the order parameter in a 
theory with purely relaxational dynamics (model A). We used a simple Metropolis 
algorithm in order to simulate the stochastic dynamics, and we employed a finite 
size scaling analysis the measure the dynamic critical exponent. We find $z = 
2.026(56)$, in agreement with predictions from the $\epsilon$ expansion, and 
earlier numerical studies. 

 We have shown that correlation functions of higher moments of the order 
parameter are governed by the same critical exponent as the two-point 
function of the magnetization. However, the relaxation time and the 
functional form of the decay depend on the power of the order parameter. 
We find that quadratic and quartic moments decay faster than the order
parameter, but the third moment decays on a time scale similar to the 
order parameter. We also studied the relaxation of powers of the order 
parameter after a quench. We thermalized the system at $T>T_c$, and determined 
the evolution using the dynamics at $T=T_c$. We find that the dynamics
involves two stages, an early time increase, followed by late time 
relaxation to the equilibrium value at $T_c$. The crossover time between
these two stages takes place at a macroscopic time $t_s\gg \Gamma^{-1}$, 
and the overall evolution does not satisfy simple dynamic scaling with 
the dynamic exponent $z$. 

 The methods discussed here can be generalized to models that are capture
more of the dynamics in a heavy ion collisions, or other physical systems
that might be of interest. In particular, there is no obstacle to considering
a conserved order parameter (model B), or the coupling to the conserved 
momentum density (model H). 

\acknowledgments
We thank Derek Teaney for useful discussions, and Chandrodoy Chattopadhyay
for pointing out an error in earlier version of this manuscript. We acknowledge
computing resources provided on Henry2, a high-performance computing cluster
operated by North Carolina State University. This work is supported by the U.S. 
Department of Energy, Office of Science, Office of Nuclear Physics through 
the Contracts DE-FG02-03ER41260 (T.S.) and  DE-SC0020081 (V.S.).

\bibliography{bib}

\end{document}